\documentclass{article}

\usepackage{arxiv}

\usepackage[utf8]{inputenc}
\usepackage[T1]{fontenc}
\usepackage[colorlinks,urlcolor=blue,linkcolor=blue,citecolor=blue]{hyperref}
\usepackage{orcidlink}
\expandafter\def\expandafter\UrlBreaks\expandafter{\UrlBreaks\do\/\do\*\do\-\do\~\do\'\do\"\do\-}
\usepackage{xurl}
\usepackage{microtype}
\usepackage{graphicx}

\title{Now's the Time: Computer Science Must Evolve to Emphasize
Software and Systems Engineering with Artificial Intelligence (AI)}

\author{
  Chandra N.~Sekharan~\orcidlink{0000-0002-5784-7400}\\
  Department of Computer Science\\
  Texas A\&M University Corpus Christi\\
  \texttt{csekharan@tamucc.edu}\\
  \And
  George K.~Thiruvathukal~\orcidlink{0000-0002-0452-5571}\\
  Department of Computer Science\\
  Loyola University Chicago\\
  \texttt{gthiruvathukal@luc.edu}
}

\begin{document}

\maketitle

\begin{abstract}
Computer science (CS) education needs to evolve to support software and artificial intelligence (AI) systems engineering, and it needs to happen now—precisely because the core intellectual contributions of CS have never been more important. We argue that traditional curricula, built around programming, data structures, and algorithms as ends in themselves, must be reframed so that these topics become foundational building blocks within a systems- and engineering-centered education. Graduates should be prepared not to compete with AI on routine coding tasks, but to design, orchestrate, verify, and own complex AI-enabled systems operating under real-world constraints. More importantly, computer science education should be geared towards preparing students for future disruptions.  The broad history of computing is marked by one disruptive technology after another, requiring us to rise to the moment instead of merely acquiescing to it.
\end{abstract}


Imagine a civil engineering graduate who has spent four years learning to lay bricks, mix concrete, and swing a hammer with precision—but has never designed a bridge, calculated load-bearing stresses, or managed a construction project where real lives and real money are at stake. Mechanical engineers do not personally weld every joint in a factory line, and electrical engineers do not hand-solder every production board. They master the fundamentals so they understand materials and processes intimately, then spend their careers architecting, integrating, verifying, optimizing trade-offs, ensuring reliability, and owning delivery of complex, real-world systems. The manual crafts are left to skilled tradespeople, or increasingly to robots, working to their specifications. Computer science education needs the same evolution.

For decades, CS degrees have drilled first-year students on Python or Java syntax, then marched them through data structures, computer organization, and algorithms as sacred rites of passage. Students grind out ``LeetCode''-style problems, implement canonical data structures, and then move on to a mix of theory, operating systems, databases, networking, and advanced mathematics. We regard this body of material as essential: students still need to recognize a data structure when they see it, understand its performance and memory trade-offs, and know why a particular algorithm or representation is appropriate in a given context. What must change is the way these topics are framed—not as ends in themselves, but as reusable design primitives and building blocks in larger systems that students will rarely, if ever, implement entirely from scratch again.

The traditional and current curriculum took inspiration from computational thinking (CT)~\cite{wing2006computational}, which, as Jeannette Wing argued, anchored the field with a powerful vocabulary for abstraction, decomposition, and algorithmic design. Properly interpreted, CT actually includes systems-oriented concerns—prefetching and caching, redundancy, damage containment, error correction, race conditions, deadlock, separation of concerns, and trade-offs between time, space, processing, and storage are all there. The real gap is that many curricula seem to have adopted a narrow interpretation of CT, stopping at algorithms and abstraction while largely downplaying the systems, uncertainty, and engineering dimensions that were present all along~\cite{DenningTedre2021}.

The result is familiar to anyone hiring: graduates who can write elegant code in isolation or in small teams, but do not always think through the more holistic considerations of shipping, scaling, securing, and maintaining systems at scale. While CS programs have changed only incrementally, enterprises and alternative providers have created parallel pathways for tech-savvy students to learn in more applied, systems-oriented settings that allow them to go straight into work. The challenge is not that CS theory, algorithms, and data structures are obsolete; it is that they are not yet being consistently taught as the conceptual ``hardware'' or ``building blocks'' of modern software and AI systems.

\section{AI Has Already Moved the Goalposts}

Generative AI has now thrown this slow drift into overdrive. Tools like Cursor, Claude, Gemini, Copilot-style assistants, and emerging agentic platforms already generate, debug, and refactor production-grade code at speeds humans cannot match for many classes of problems. We acknowledge that these tools are far from perfect, but their rate of improvement is unmistakable, and they  will continue to improve at a rapid pace. We joke with our students that if you sleep for too long, a new or improved AI model might be there when you wake up. The irony is that the very systems thinking we are arguing for is what made modern AI possible in the first place. Large-scale AI is not the product of a single clever algorithm; it is the outcome of systems thinking at massive scale—breakthroughs in distributed systems, data pipelines, and hardware integration, layered with careful engineering in pretraining, fine-tuning, and reinforcement learning from human feedback (RLHF) to transform a stochastic next-token predictor into one of the most useful artifacts in contemporary tooling. This massive scale itself is beyond the capabilities of most institutions and is now the province of just a handful of companies that were pioneers in cloud computing.

The industry is already reorganizing around this reality. Gartner, for example, projects that within a few years the overwhelming majority of enterprise software engineers will be using AI code assistants on a daily basis~\cite{gartner2025toptrends}, with roles shifting from ``implementation'' to ``system design and orchestration''. McKinsey \& Company's 2025 work on AI in software development finds that the highest-performing organizations see sizeable gains in productivity, time to market, customer experience, and software quality—on the order of tens of percentage points—by embedding AI across the entire product-development lifecycle and redefining roles and practices accordingly~\cite{mckinsey2025unlockingai}. Early-career roles are already feeling the pressure: recent work from the Stanford Digital Economy Lab reports that since the widespread adoption of generative AI, early-career workers (ages 22–25) in the most AI-exposed occupations, including software developers, have experienced a considerable relative decline in employment, with adjustments occurring primarily through reduced hiring rather than wage cuts~\cite{brynjolfsson2025canaries}. Routine programming is turning into a commodity, not because software is going away, but because the nature of software work is changing.

We have been here before, at smaller scales. Mainframe computers and personal computers (PCs) were supposed to wipe out office work; instead, they reshaped it. The World Wide Web was going to destroy ``brick and mortar''; instead, it rewired commerce; ``brick and mortar'' remains and likely will do so in an AI-enabled world. Wikipedia was feared as a flattening force that would replace experts; instead, it became part of the foundational infrastructure—and training data—for today's AI, which basically relies on a snapshot of the Internet as a crucial aspect of its training. Each time, computing reinvented itself as a distributed, networked, systems discipline. The internet's evolution into the web and then into cloud computing is, at heart, a story of systems and scale. This latest AI wave is consistent with that pattern but more intense, and it demands that we raise the level of abstraction and awareness in what we teach beyond the basic coding tasks that were already influenced by information available from sites like Stack Overflow and Chegg long before large language models (LLMs) arrived. Suffice it to say, not only has the time come but it may have arrived before we were ready.

This is part of what Peter Denning has been arguing for years. His ``Great Principles'' framework~\cite{denning2015great} positions computing not as a narrow technology or coding craft but as a science of information processes governed by principles like communication, coordination, and design. In that framing, today's AI systems and the infrastructures around them are not anomalies; they are archetypal examples of what computing is supposed to study and build, drawing deeply on algorithms, data structures, theory, and formal methods even when those are no longer implemented ``by hand'' line by line.

\section{Re-centering CS on Systems and Engineering}

Academic CS programs now face a stark question: are they preparing students to compete with AI on tasks that AI increasingly does better and faster, or to direct and integrate AI as master systems engineers direct construction crews, or as composers orchestrate virtuoso musicians into a coherent performance? The answer cannot be to abandon programming or devalue core CS material. Instead, we should reposition programming, algorithms, and data structures to the kind of foundational status that materials science and shop practice hold in traditional engineering. Students still need to prototype, read, debug, and verify code—especially AI-generated code—early and often, because you cannot specify, critique, or audit what you do not understand.

Future systems-enabled computer science graduates should know, for example, when a hash table, balanced tree, or graph representation is being used under the hood, what its asymptotic and constant-factor costs are, and how those choices affect latency, memory use, and fault behavior in a distributed setting. What they should not need to do, as a core professional skill, is reimplement these structures from first principles every time in the language of the day; instead, they must be able to treat them as reliable components whose properties they understand and can reason about in larger designs and trade-offs. In this sense, the classic CS topics become the grammar of a systems language: still indispensable, but no longer the whole story.

Beyond that foundation, the center of gravity in the computer science degree should move toward the kinds of work that actually create value in the AI era. Systems thinking and architecture should become front and center: problem framing, abstraction, and the design of resilient, modular systems must come before the incidental syntax of any one language. Coordination and agentic orchestration should become core topics, teaching students how to manage complex ``conversations'' among AI agents, services, and humans to maintain goal alignment and distributed consistency across an evolving ecosystem. Design science must expand beyond traditional user interface and user experience (UI/UX) design to consider how systems are adopted in practice, how automation reshapes work, and how to design software that humans can trust, govern, and integrate into their workflows. Speaking specifically to UI/UX, AI and Agents have,  ironically, returned the user interface to a simple textual one, not altogether removed from the basic command-line prompt.

Domain-driven integration belongs explicitly in the curriculum: gluing together services, data pipelines, and AI components into coherent wholes is where much of the hard work now lives. Reliability and verification engineering must be taught not as niche advanced electives but as mainstream concerns: chaos testing, observability, and rigorous validation of LLM outputs, using self-consistency checks, grounding to authoritative sources, hallucination detection, and automated scoring, are now table stakes for any serious AI-enabled system. Infrastructure, delivery, and cost engineering should treat cloud bills, latency, and uptime as first-class constraints, just as civil engineers treat material cost and structural integrity. Ethics, security, and professional ownership need to be woven throughout, not bolted on; the ethical, legal, and security boundaries around AI systems are being redrawn in real time, and ``owning what you built'' is fast becoming a non-negotiable expectation.

A capstone experience in this model looks very different from a traditional term project. Students would work with real stakeholders and users, operate within explicit cost budgets (e.g., cloud credits), and aim for real impact through production deployment, open-source contributions, or meaningful prototypes. They would be expected to deliver architecture documents, continuous integration and continuous delivery (CI/CD) pipelines, observability and reliability setups, cost dashboards, security and ethics justifications, and post-mortem reflections. Throughout, the ``classic'' CS content such as data structures, algorithms, operating systems, networking, and databases would show up repeatedly as the conceptual toolkit used to justify design choices and analyze system behavior, not as conceptual silos.

Equally important is preparing students to reason about system dynamics when the state-transition rules are no longer written in code. LLM-based agents maintain state, sense their environment, and decide next actions through semantically driven, stochastic processes that emerge from prompts and model behavior. The resulting state trajectories are not easily analyzable or verifiable using classical state machine methods alone. A developer may voluntarily migrate to a newer model seeking better performance and/or lower cost or may be forced to migrate when the current version reaches end-of-life. In either case, the API contract is typically preserved: the same endpoints, the same parameters, the same response schema. What is not preserved, and what the developer has no reliable method to assess, is behavioral equivalence — the new model may interpret prompts differently, resolve edge cases with different biases, or shift classification thresholds in ways that conventional test suites, designed around structural correctness rather than semantic consistency, will not catch.

Students must therefore learn to design for underlying instability through behavioral regression testing, shadow deployment, and architectures that treat the LLM as a swappable component with explicit behavioral expectations above the API layer. Industry practitioners who build and maintain systems at scale may need to be involved as  active partners in shaping this curriculum while preserving academic independence.

\section{Elevating, Not Diluting, Computer Science}

None of what we argue here amounts to ``dumbing down'' computer science; nor is it a repudiation of CS as a discipline.  It is the opposite. It is treating software, and now AI-orchestrated software, as an engineered artifact that must be designed, stressed, deployed, broken, and understood under real constraints, using the full depth of CS theory and practice. The intellectual core of the discipline remains; abstraction, trade-offs, rigorous reasoning, and mathematical foundations are not going anywhere. They are simply being relocated to where they matter most: in the design, analysis, and stewardship of complex, AI-augmented systems that deliver value at scale under uncertainty and societal scrutiny.

Departments that make this leap will graduate students who are immediately valuable to future employers. These graduates will walk into industry ready to work on production platforms, direct fleets of AI agents, verify mission-critical outputs, and ship systems that do more than merely run—they will survive and improve under real load, cost pressure, security threats, and regulatory oversight. Programs have an opportunity right now to lead in reimagining curricula in light of what AI is commoditizing on a near-daily basis, while explicitly honoring and leveraging the deep CS knowledge that made these tools possible in the first place. One of the most exciting aspects of computer science has always been its habit of inflicting upon itself, in the best possible sense, a newly evolved world every few decades. We are in the middle of one of those moments. And it will not be the last such moment.

\section{Conclusion}

The bottom line is that a new CS curriculum is not about copying traditional engineering syllabi; it is about co-opting the engineering mindset while preserving the discipline's intellectual core. Treat software like a machine: design it, stress it, deploy it, break it, and learn from it. Your ``bridge'' might be an autonomous drone swarm, a fraud detection platform, or an AI-native workflow engine that coordinates dozens of agents. The question for every computer science program is now unavoidable: how will you lead this evolution for the future of the discipline, for the careers of the next generation of students who trust you with their education, and for the continued vitality of computer science itself?

\bibliographystyle{unsrt}





\end{document}